\documentstyle[emulateapj,psfig,apjfonts]{article}
\submitted{Received 2005 April 14; Accepted: 2005 June 8}

\def\gs{\mathrel{\raise0.35ex\hbox{$\scriptstyle >$}\kern-0.6em
\lower0.40ex\hbox{{$\scriptstyle \sim$}}}}
\def\ls{\mathrel{\raise0.35ex\hbox{$\scriptstyle <$}\kern-0.6em
\lower0.40ex\hbox{{$\scriptstyle \sim$}}}}

\newenvironment{inlinefigure}{%
\def\@captype{figure}%
\noindent\begin{minipage}{0.999\linewidth}\small}
{\end{minipage}\smallskip}
\makeatother

\lefthead{Smail, Smith \& Ivison}
\righthead{The Intrinsic Properties of SMM\,J14011+0252}

\begin{document}

\title{The Intrinsic Properties of SMM\,J14011+0252}

\author{
Ian Smail,\altaffilmark{1} G.\,P.\ Smith\altaffilmark{2}
\& R.\,J. Ivison\altaffilmark{3,4} }

\altaffiltext{1}{Institute for Computational Cosmology, University of Durham, South Road,
        Durham DH1 3LE UK}
\altaffiltext{2}{California Institute of
Technology, Department of Astronomy, MC 105-24, Pasadena, CA 91125}
\altaffiltext{3}{Astronomy Technology Centre, Royal Observatory,
Blackford Hill, Edinburgh EH9 3HJ}
\altaffiltext{4}{Institute for Astronomy, University of Edinburgh,
Blackford Hill, Edinburgh EH9 3HJ}

\setcounter{footnote}{4}

\begin{abstract}
We discuss the properties of the bright submillimeter source
SMM\,J14011+0252 at $z=2.56$ which lies behind the central regions of
the $z=0.25$ lensing cluster A\,1835.  This system has a complex
optical morphology consisting of at least five separate components.  We
reassess the extensive multiwavelength observations of this system and
find strong support for the suggestion that one of these five
components represents a foreground galaxy.  The spectral and
morphological properties of the foreground galaxy indicate that it is a
low-luminosity, passive early-type disk member of the A\,1835
cluster.  We estimate the likely properties of the dark matter halo of
this galaxy from its stellar distribution.  Based on these estimates we
suggest that, contrary to earlier claims, this foreground galaxy is
unlikely to significantly magnify the background submillimeter source.
Thus SMM\,J14011+0252 probably represents an intrinsically luminous
submillimeter galaxy.
\end{abstract}


\section{Introduction}

%
%
\begin{figure*}[tbh]
\centerline{\psfig{file=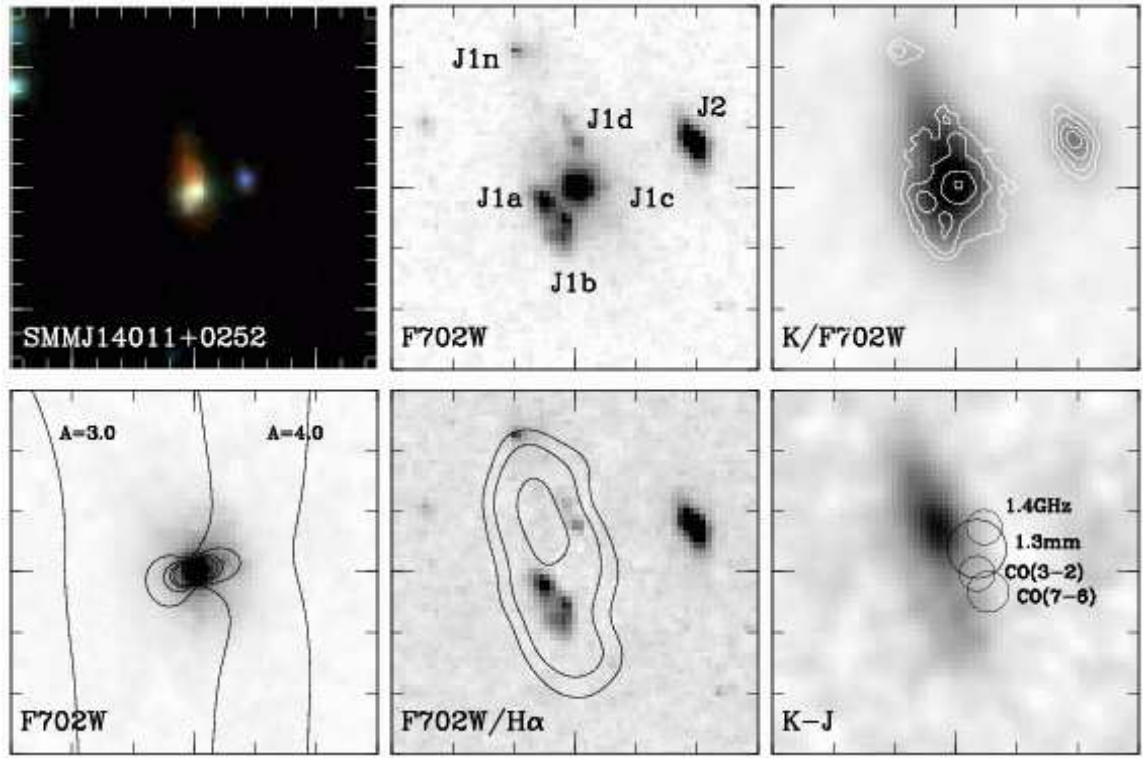,width=6.0in,angle=270}}
\caption{\small 
Six different views of the SMM\,J14011 system. The top row comprises
(starting at the left): a true-color image of the galaxy produced from
the R$_{702}$JK imaging; the WFPC2 R$_{702}$ image with the various
components labelled (following the naming scheme of M05); the Keck
K-band image from Frayer et al.\ (2004) with the R$_{702}$ image
overlayed as a contour.  The panels in the bottom row show (again
starting from the left): the R$_{702}$ image of the component J1c,
proposed to be a foreground cluster galaxy, overlayed on this is a
contour plot showing the estimated amplification factors from our lens
model; the various optical components of the background galaxies, with
the continuum-corrected H$\alpha$ image from T04 shown as a contour; an
image of the emission in the K-band obtained by subtracting off the
scaled J-band emission from J1c.  Each of the panels is
6$''\times$6$''$ with North to the top and East to the left, apart from
the true color image which is 15$''\times$15$''$. The data shown in the
various panels comes from I01, Frayer et al.\ (2004), and T04.
[Resolution degraded due to file size limit]
}
\end{figure*}

SMM\,J14011+0252 (hereafter SMM\,J14011) is a bright submillimeter
(submm) source identified in the field of the cluster lens, A\,1835, by
Ivison et al.\ (2000, I00) (see also Smail et al.\ 2002).  The submm source
has a radio counterpart, allowing the position of the submm emission to
be pin-pointed to a close pair of blue galaxies.  The redshift of the
system was measured as $z\sim 2.56$ from a near-infrared spectrum by
I00, and both components were confirmed to lie at the
same redshift from optical spectroscopy by Barger et al.\ (1999).
Molecular CO emission was subsequently identified at this redshift by
Frayer et al.\ (1999).  These observations appeared to show that the
$z=2.56$ system is a massive, gas-rich far-infrared luminous source,
only the second such system identified from submm surveys (e.g.\ Ivison
et al.\ 1998).

SMM\,J14011 has become a archetype for submm galaxies, being studied
extensively by a number of authors (Ivison et al.\ 2001, I01; Downes \&
Solomon 2003, DS03; Tecza et al.\ 2004, T04; Frayer et al.\ 2004;
Swinbank et al.\ 2004; Motohara et al.\ 2005, M05; Carilli et al.\
2005).  From the outset it had been assumed that SMM\,J14011 was
gravitationally amplified by the potential of the foreground cluster,
by a factor of $\mu \sim 2.8$--$3 \times$ (I00; Frayer et al.\ 1999).
However, DS03 suggested that in fact the submm source is highly
amplified by a superimposed, foreground galaxy -- resulting in J1 and
J2 being two images of a single background source, with an
amplification of a factor of $\mu \sim 25$--30 (see also M05).  This
scenario results in very different conclusions about the intrinsic
properties of SMM\,J14011 and would mean that the source is not
representative of the mJy-submm population, which are the subject of
much research.

In this paper we reassess the available multiwavelength information on
SMM\,J14011 to determine the likely influence of gravitational lensing
on the measured properties of this galaxy.  We assume a cosmology with
$\Omega_m=0.27$, $\Omega_\Lambda=0.73$ and
$H_o=71$\,km\,s$^{-1}$\,Mpc$^{-1}$, giving an angular scale of
3.88\,kpc and 8.15\,kpc per arcsec.\ at $z=0.25$ and $z=2.56$
respectively.

\section{Analysis and Modelling}

We show a variety of different views of SMM\,J14011 in Fig.~1 and mark
on the main components following the naming scheme of M05.  From
ground-based optical imaging it is clear that the system is made up of
two main components: J1 and J2 (I00).  J1 and J2 are both
relatively blue, compared to the bright early-type galaxies in the
foreground cluster (I00).  However, higher resolution optical imaging
from {\it Hubble Space Telescope} (I01) identifies several knots within
J1 -- J1a/J1b/J1d in Fig.~1, which appear superimposed on the
relatively smooth and regular underlying component J1c.  While
high-resolution ground-based near-infrared imaging highlights a
low-surface brightness, very red extension to the north of J1, which
terminates in a faint knot, J1n (I01; Frayer et al.\ 2004).  The
morphology of the redshifted H$\alpha$ emission from the $z=2.56$
galaxy has been mapped using both narrow-band imaging (Swinbank et al.\
2004) and integral field spectroscopy (T04).  These observations show
that the H$\alpha$ emission is seen from the knots J1a/J1b/J1d and the
$\sim 2.5''$-long, very red extension out to J1n (Fig.~1) -- but not
from J1c.

The exact position of the submm, molecular gas and radio emission
relative to the optical imaging remains somewhat uncertain.  Aligning
the {\it HST} image to the radio reference frame using the five radio
sources with bright optical counterparts (but not SMM\,J14011) yields a
position for J1c of 14\,01\,04.96 +02\,52\,24.1 (J2000) with an
uncertainty of $\sim 0.3''$ (consistent with the published position
from I01).  We plot the centroids and 1-$\sigma$ error-circles for the
CO(4--3) and CO(7--6) emission from DS03 and our 1.4-GHz centroid on
Fig.~1.  Within the relative uncertainties the radio position could be
associated with J1d or the near-infrared peak next to it, although the
CO peaks are both somewhat south of this position.

~\bigskip

\subsection{The true nature of J1c}

A detailed inspection of the optical spectrum of J1 from Barger et al.\
(1999), suggests that it comprises a mix of light from a UV-bright
starburst at $z=2.56$ and a foreground galaxy as proposed by DS03.  To
isolate the light from the foreground galaxy we scale and subtract the
spectrum of a high-redshift starburst (J2) from the spectrum of J1.
The scaling factor is chosen so that the absorption features around
$\sim 4000$\AA\ are not highly negative in the resulting spectrum.  The
difference spectrum is shown in Fig.~2 and exhibits a number of
absorption features which are identifiable as Balmer H$\theta$
[4743.3\AA\ observed wavelength], Ca H [4912.0\AA], Ca K [4956.8\AA],
H$\delta$ [5126.7\AA], G-band [5375.8\AA] and H$\beta$ [6069.3\AA] at
$z=0.2489\pm 0.0004$.  Thus it appears that there is light from a
foreground member galaxy of the A\,1835 cluster mixed with that from
the $z=2.56$ system in our spectroscopic slit.  The identification of
this foreground component is hampered by an unfortunate coincidence
that some of the absorption lines in the $z=0.25$ galaxy can be
interpreted as UV features at the redshift of the background source,
e.g.\ C{\sc ii}1335 corresponds to 4743.3\AA\ and Si{\sc iv}1394 falls
close to 4956.8\AA.

The lack of any detectable emission lines from the $z=0.25$ component
in the J1 spectrum (e.g.\ [O{\sc ii}]\,3727, [O{\sc iii}]\,5007), along
with the apparent strength of the Balmer absorption lines (EW(H$\delta) \sim
3$\AA) suggest that the superimposed galaxy could either be a passive
galaxy or potentially a post-starburst system (Poggianti et al.\ 1999).
However, we caution that the strength of these features is strongly
dependent on the relative contributions from the foreground and
background components in the spectrum, which is not well-determined.

Thus we confirm the suggestion by DS03 that one component of J1 is a
foreground galaxy.  Looking at the morphologies of the various
components in Fig.~1, it seems most likely that J1c corresponds to the
$z=0.25$ galaxy.

To gain a clearer view of the morphology and luminosity of J1c we take
advantage of its apparent regularity and symmetry to remove the
superimposed knots of emission.  We rotate the {\it HST} image of
SMM\,J14011 by 180\,degrees and subtract it from itself -- producing an
image of just the knots of emission: J1a/J1b/J1d and J2.  These can
then in turn be subtracted from SMM\,J14011 to leave just the emission
from J1c.  This shows the restframe $V$-band structure of the galaxy --
with a smooth, circular morphology indicative of a spheroid or face-on
early-type disk.  The measured ellipticity of J1c is $\epsilon=0.03\pm
0.01$ (at PA=133\,deg.) and its total magnitude (within a
6$''$-diameter aperture) is $R_{702}=21.27\pm 0.03$ (our best estimate
of the $K$-band magnitude of J1c is $K=18.44\pm 0.10$, giving a color
of $(R-K)\sim 2.9$, similar to that expected for a dwarf spheroid or
early-type disk cluster member, Smith et al.\ 2002).  Using {\sc
gim-2d} (Simard et al.\ 2002) we model the 2-D light distribution in
J1c and obtain a best fit model ($\chi^2=1.3$) with a half-light radius
of $r_{hl}=0.40\pm 0.05$ or $\sim 1.6$\,kpc and a bulge-to-disk ratio
of $0.30\pm 0.05$, suggesting the galaxy is an early-type disk (Sab).

The apparent magnitude of J1c corresponds to an absolute magnitude of
$M_r\sim -19.2$ at $z=0.25$, or roughly 0.1$L^\ast$.  Assuming the
galaxy follows the Faber-Jackson relation measured in moderate redshift
clusters (e.g.\ Zeigler et al.\ 2001), we would expect a velocity
dispersion for the galaxy of $\ls 55\pm15$\,km\,s$^{-1}$ given its
absolute magnitude (the error is the 1-$\sigma$ scatter of galaxies
around the Faber-Jackson relation).  We quote this value as a limit due
to the possibility that the mass-to-light ratio of J1c is higher than
typical passive galaxies in the cluster if it has suffered recent
star-formation.  The low velocity dispersion we infer for this galaxy
is consistent with the small half-light radius we measure if it follows
the scaling ratios for early-type cluster galaxies at $z\sim 0.2$.
Thus we conclude that J1c is a passive, early-type dwarf disk galaxy
member of A\,1835.

%
%
\begin{inlinefigure}\vspace{6pt}
\psfig{figure=f2.ps,width=3.5in,angle=270}
\vspace{6pt}
\noindent{\small {Fig.~2. --- }
The spectrum of J1c obtained by subtracting a $z=2.56$ Lyman-break
galaxy (a scaled version of the J2 spectrum) from the published J1
spectrum of Barger et al.\ (1999).  The spectrum shown in bold is the
smoothed difference spectrum, while the upper and lower traces show the
spectra of component J1 and J2 respectively (offset for clarity).  The
difference spectrum represents the light from the $z=0.25$ galaxy, J1c.
We mark some of the stronger spectral features visible in the
foreground galaxy spectrum and we note that J1c contributes roughly
half the light observed at 5000\AA\ in the spectrograph slit. }
\end{inlinefigure}

We also use take advantage of the very different $(J-K)$ colors of J1c and
the red-extension, J1n (I01; Frayer et al.\ 2004) to scale the J-band
emission of J1c and subtract it from the K-band image to leave a clear
view of the K-band morphology of the high-redshift galaxy, Fig.~1.
This unsurprisingly matches the H$\alpha$ morphology from T04, given
the strong contribution from the H$\alpha$ line in the K-band (I00;
T04; Swinbank et al.\ 2004; M05).  However, we can use the good
resolution of the Keck imaging (0.4$''$ FWHM) to show that this
emission is resolved in both axes with a seeing-corrected size of
$1.8''\times 0.6''$.

\subsection{The strong lensing interpretation of SMM\,J14011}

Two strong lensing models have been proposed for SMM\,J14011 based on
the identification of various components of the system as multiple
images of single background sources.  DS03 discuss a toy model which
seeks to explain J1a/J1b/J1n and J2 as four images of a single
background source.  Although they do not construct a rigorous model
they suggest that a potential well associated with J1c with a velocity
dispersion of $\sim 200$\,km\,s$^{-1}$, along with a contribution from
the cluster mass distribution, could fit the four proposed images.

As shown by Swinbank et al.\ (2004) and T04, the spectroscopic
information available for J1a/J1b/J1n and J2 indicates different
redshifts for the two systems from their H$\alpha$ emission (e.g.\
Fig.~1).  This rules out J2 as a counter image and hence allows us to
discard the qualitative model suggested by DS03.

M05 also fit a strong-lensing model to components of J1 -- this time
J1a, J1b and J1d -- using the {\sc gravlens} model of Keeton (2001).
Their model includes an external shear field arising from the cluster
potential (constrained by the mass model of Schmidt et al.\ 2001) and
is capable of reproducing the positions of the three proposed images
using a highly-elliptical ($\epsilon=0.67\pm0.09$,
PA\,$=53\pm10$\,deg.) potential well centered on J1c with a velocity
dispersion of $123^{+10}_{-14}$\,km\,s$^{-1}$.  

The obvious problem with the M05's lens model is that the
parameters of the potential well of J1c differ substantially from those
traced by the light distribution within this galaxy on similar scales.
The ellipticity, position angle and likely velocity dispersion of the
halo are all strongly at odds with those measured or expected for J1c
($\epsilon=0.03\pm0.01$, PA\,$=133\pm5$\,deg.\ and $\sigma\ls 55\pm
15$\,km\,s$^{-1}$) if mass traces light on $\sim 10$\,kpc scales within
this galaxy in a similar manner to that found from modelling strong and weak
lensing by individual galaxies (Treu \& Koopman 2004; Kochanek et al.\
2000; Natarajan et al.\ 1998).  In particular, we note that the
velocity dispersion required in this model would correspond to a galaxy
with an absolute magnitude of $M_r\sim -21$, some $6\times$ brighter
than J1c (assuming the $z\sim 0.2$ Faber-Jackson relation from Zeigler
et al.\ 2001).

In addition, we find that the predicted flux ratios between J1a/J1b of
11:13 are at odds with the measured 1.0$''$ aperture magnitudes from
the {\it HST} image of $R_{702}=23.54$ and 23.91 (with uncertainties of
$\pm 0.05$, J1d has $R_{702}\sim 24.2$ in the same aperture).  These
yield a flux ratio between J1a and J1b of $0.71\pm 0.07$, compared to
the predicted ratio of 1.18.

All of these discrepencies arise because of the incorrect
identification of J1a/J1b/J1d as three images of a single background
source.  Looking at Fig.~1, it is clear that there is no compelling
similarity between the morphologies of J1a, J1b and J1d: J1b is well
resolved, with a FWHM of 0.57$''$, while J1a is more compact
(FWHM=0.30$''$) and J1d is too faint to allow us to measure a reliable
size.

\subsection{Lens models}

The observed properties of the various components of SMMJ14011 do not
appear to {\it require} a strong-lensing interpretation.  However
gravitational magnification due to the known mass concentrations along
our line-of-sight to SMMJ14011, i.e.\ J1c and A\,1835, inevitably
modify its observed properties.  We quantify this effect, first
considering the magnifying power of J1c alone.  Adopting
55\,km\,s$^{-1}$ as the line-of-sight stellar velocity dispersion of
J1c, and a simple singular isothermal model ($\sigma_{\rm
1D}=\sqrt{4/3}\times\sim 55$\,km\,s$^{-1} =64$\,km\,s$^{-1}$), the
Einstein radius of J1c is just $\sim0.1''$.  In contrast, the observed
separation of J1a/J1b from J1c is $0.55''$.  To achieve an Einstein
radius comparable with the observed separation, J1c would require a
velocity dispersion of $\sigma_{\rm 1D}\sim150$\,km\,s$^{-1}$ (as
concluded by M05), implying an anomolously high mass-to-light ratio.

We now consider to what extent the contribution from the cluster
potential can ameliorate this situation using the {\sc lenstool} lens
modelling code (Kneib et al.\ 1993, 1996; Smith et al.\ 2005, S05).
The Schmidt et al.\ (2001) mass model for the cluster adopted by M05
uses ground-based optical photometry of selected arcs and the cluster's
X-ray emission to constrain a model of the cluster mass distribution.
However, {\it HST} imaging now reveals one of the multiple-image
systems (B and B$'$) employed in their modelling by Schmidt et al.\ to
be unreliable (S05).  We therefore instead use the gravitational lens
model developed using {\it HST} WFPC2 imaging by S05 to describe the
mass distribution within the cluster.  We refer the interested reader
to S05 for full details of the model.  The key features relevant to
this study are (i) despite the presence of several gravitational arcs
in A\,1835, none of them have been spectroscopically identified to date
(S05), hence the model is constrained by the weak shear signal detected
in the outskirts of the WFPC2 frame, (ii) this weak shear signal was
calibrated to 10\% precision in projected mass using
spectroscopically-confirmed strong-lensing clusters at the same
redshift as A\,1835, (iii) the family of acceptable models (defined by
$\Delta\chi^2\le1$) bracket Schmidt et al.'s models.

Taking our best-fit mass model for A\,1835 we can now add a mass component at
the position of J1c with ellipticity, position angle and velocity
dispersion matching our estimated values (\S2.2).  The surface mass
density of the combined A\,1835+J1c lens is sub-critical at the
position of J1a/J1b/J1d, i.e.\ the combined lens is not capable of
multiple-imaging at the location of the submm emission.  The same is
true of models at the high and low-mass extremes allowed by the
weak-shear data and the observational constraints on J1c.  In summary,
we estimate a conservative 1-$\sigma$ range for the amplification of
the various components of: J1a $\mu=2.8$--4.9; J1b $\mu=2.7$--4.1; J1d
$\mu=2.7$--4.0; J1n $\mu=2.7$--3.9; J2 $\mu=3.0$--5.0; the
uncertainties are driven by the statistical error from the weak-shear
constraints on A\,1835.  We therefore conclude that it is unlikely that
SMM\,J14011 is strongly lensed, instead the different observed
components are likely magnified by factors of $\mu\sim3$--5.

\section{Discussion}

Our analysis confirms the suggestion of DS03 that a
component of the complex submm source, SMM\,J14011, is actually a
foreground galaxy.  However, we do not support their suggestion
that the background far-infrared luminous source is multiply-imaged and
hence highly-amplified by the foreground galaxy (in combination with
the A\,1835 cluster).  There is clear spectroscopic evidence which
rejects their proposed identification for the multiple images.
Similarly, we have used the measured properties of the foreground
galaxy, J1c, to reject a second lensing configuration suggested by
M05.  Both the observed ellipticity and estimated
mass of the halo of J1c are very different from those required by 
M05's model, while the flux ratios and morphologies of the
proposed counter-images also do not agree with their predictions.

We suggest that J1c is probably a passive dwarf member of the A\,1835
galaxy cluster.  Our best estimates of the likely velocity dispersion
of this galaxy, $\sigma \ls 55\pm 15$\,km\,s$^{-1}$, indicate it is
incapable of producing multiples images on the observed angular scales
of J1a/J1b/J1d.  Instead, we estimate a modest boosting of the
amplification of these images, over that provided by the cluster alone.
Assuming that the submm emission broadly traces the H$\alpha$ emission
mapped by T04 gives a median amplification averaged over the source of
$\mu\sim 3.5\pm 0.5$, slightly higher than the original estimate
assumed by I00 based purely on the lensing influence of the cluster
potential.  Thus SMM\,J14011 has an intrinsic submm flux of $3.5\pm
0.5$\,mJy at 850\,$\mu$m and a far-infrared luminosity of $L_{FIR}\sim
4\times 10^{12}L_\odot$. SMM\,J14011 is an intrinsically luminous galaxy
(Smail et al.\ 2002).

To study the structure of the background galaxy in more detail we can
correct its observed shape (as displayed in the $K-J$ image in Fig.~1)
for the distortion produced by the lens using our model.  This
indicates that the source is likely to have an intrinsic FWHM of $\sim
0.5''$ (4\,kpc) and has a relatively circular morphology in the
restframe optical (perhaps corresponding to a face-on orientation,
which would help explain the relatively narrow CO line width for the
system, DS03).  The knots, J1a/J1b/J1d/J1n, visible in the {\it HST}
imaging appear to be UV-bright clumps lying in a $\sim 10$\,kpc region
corresponding to the restframe optical extent of the galaxy and are
most likely relatively unobscured star-forming regions within the
galaxy.  J2 represents a UV-bright companion with a 
separation of just $\sim 20$\,kpc in projection.  The close proximity
of J2 to J1 suggests that their dynamical interaction may be
responsible for the intense starburst currently underway in J1.  The size and
clumpy/multi-component morphology of SMM\,J14011 is thus very similar
to that of typical submm galaxies with submm fluxes of $\sim5$\,mJy
(Chapman et al.\ 2004; Smail et al.\ 2004).

We conclude that the intrinsic nature of SMM\,J14011 is much as
originally stated by I00, at least regarding the
long-wavelength properties (where J1c does not contribute).  However,
the identification of J1c as a foreground contaminant with only a
weak lensing contribution does alter the conclusions of DS03 and M05
who both adopted strong-lensing models to correct the observed
properties of SMM\,J14011.  Using our prefered lens model, we
reinterpret the apparent CO(7--6) size of SMM\,J14011 from DS03, who
constrain the source to be $2.3''\times \ls 0.8''$ (corrected for the
beam), which corrected for lens amplification indicates an intrinsic
CO(7--6) FWHM of $\sim 6$\,kpc.  This makes the CO emission in
SMM\,J14011 comparable in size to the few SMGs studied at
high-resolution with IRAM (Genzel et al.\ 2003; Tacconi et al.\ 2005).
The discussion in DS03 then suggests that the source must plausibly
consist of a series of compact knots distributed within the beam if the
brightness temperature is not going to be too high.

In the case of M05, the main conclusion which changes as a result of
our lens model is the estimated stellar mass of the system, which
instead of being $\ls 10^9$M$_\odot$ (adopting an amplification of $\mu
\sim 30$), should be closer to $\ls 10^{10}$M$_\odot$, comparable to the
gas mass and dynamical mass limits derived from the CO line width
(Frayer et al.\ 1999).  Again this makes the restframe
optically-derived stellar mass of SMM\,J14011 very similar to that of
comparably luminous submm galaxies in the field (Smail et al.\ 2004),
although we caution that these stellar mass estimates are highly
uncertain due to the large extinction correction and sensitivity to the
adopted ages and hence mass-to-light ratios of the stellar populations
(Borys et al.\ 2005).

We note that the identification of J1c as a foreground galaxy will also
alter the model of the near-infrared spectral energy distribution of
this system derived from high-quality 2-dimensional spectroscopy by
T04.  They attempt to reproduce an apparent spectral break between the
$J$- and $H$-bands with the Balmer discontinuity in a young stellar
population and derive a rough age of 200\,Myrs and a reddening of
$A_V\sim0.7$ from this.  From the observed $K$-band magnitude of J1c, we
estimate that roughly half of the light in the $K$-band spectrum of T04
comes from the foreground galaxy and adopting reasonable
near-infrared colors of $(J_{AB}-H_{AB})\sim 0.3$ and
$(H_{AB}-K_{AB})\sim 0.2$ for this galaxy (assuming an unevolved
early-type spectral energy distribution at $z=0.25$), that the true
spectral break is likely to be stronger when the contribution from J1c
is removed.  This would likely decrease both the age and the reddening
derived by T04 -- bringing these closer to the values determined by
M05.  However, we stress that T04's analysis of the baryonic mass of
this system relies on the spectral rather than photometric properties
of this system and hence is insensitive to the exact properties of J1c.

\section{Conclusions}

Our analysis of the spectral properties of J1c, a
proposed component of the $z=2.56$ submm source SMM\,J14011, confirms
that it is in fact a dwarf early-type member of the foreground cluster.
However, our detailed modelling of the gravitational lensing contribution from
J1c on the background submm source indicates that it is unlikely to
significantly increase the amplification of the source over that from
the cluster potential alone.  We estimate conservative limits on the
amplification of the various UV/near-infrared components of SMM\,J14011
in the range $\mu \sim 3$--5.  We conclude that SMM\,J14011 remains an
intrinsically very luminous galaxy with properties that appear similar
to comparably luminous systems now being studied in greater numbers
at high redshifts.

\acknowledgments
We thank Dave Frayer, Jean-Paul Kneib, Nicole Nesvadba and Mark
Swinbank for help and Andrew Baker, Dennis Downes and Kentaro Motoharo
for useful discussions. We thank an anonymous referee for a
report which helped improve the presentation of this paper.
IRS acknowledges support from the Royal Society.

\end{document}